\documentclass{rec12}    

\usepackage{graphicx}
\usepackage{times}
\usepackage{color,amsbsy}

\begin{document}
\begin{opening}
\title{ Variance-based sensitivity analysis\\ 
in the presence of correlated input variables}
\author{Thomas \surname{Most}}
\runningauthor{Thomas Most} \runningtitle{Variance-based sensitivity analysis in the presence of correlated input variables}
\institute{Dynardo -- Dynamic Software and Engineering GmbH, Weimar, Germany\\ thomas.most@dynardo.de}
\date{}

\begin{abstract} In this paper we propose an extension of the classical Sobol' 
estimator for the estimation of variance based sensitivity indices. 
The approach assumes a linear correlation model between the input variables
which is used to decompose
the contribution of an input variable into a correlated and an uncorrelated part.
This method provides sampling matrices following the
original joint probability distribution which are used 
directly to compute the model output without any assumptions or
approximations of the model response function.
\end{abstract}

\end{opening}

\section{Introduction}
In the assessment of computational engineering models sensitivity analysis has played an increasing role during the last
two decades. For probabilistic models variance based sensitivity methods are very common.
Based on the work of Sobol' \shortcite{Sobol1993} first order indices have been extended to
total indices capturing higher order coupling effects between the input variables \shortcite{Homma1996}.
Based on this early work a large number of extensions and improvements have been developed.
A precise overview of existing variance based methods and their background is given in \shortcite{Saltelli_2008_Book}.
As stated in \shortcite{Saltelli2010} the total effect sensitivity index has become a 
prestigious measure to quantify higher order effects of input variables which is necessary
for model simplification by variable fixing. 
However, the basic formulation of the variance decomposition and the common estimation procedures require so far 
independent input variables.

In practical applications often input variables are correlated, therefore, 
an increased interest in extending the classical methods for correlated inputs
is marked in the recent years.
In several investigations, 
assumptions about the model output or approximations up to a certain degree are necessary \shortcite{Oakley2004}, \shortcite{Xu2008}, 
\shortcite{DaVeiga2009}. Complete model independent approaches for the estimation of total effect indices are not
available so far. In \shortcite{Jaques2006} classical estimators have been used by grouping the correlated inputs
which results in independent groups of parameters. However, even this approach is not useful for more complex models where 
almost all variables are coupled with each other. 

In this paper we propose an extension of the classical Sobol' estimator. 
The basics of this method have been developed in 
\shortcite{Most_Saltelli_2010}, where a variable decomposition similar to \shortcite{Xu2008} was realized. 
The general idea has been adapted in \shortcite{Mara2012}
for normally distributed random variables.
In our
approach we assume a linear correlation model between the input variables, 
which is used to decompose
the samples into a correlated and an uncorrelated part with respect to a certain variable.
This was realized in \shortcite{Mara2012} by an orthogonalization schemes which may lead to non-unique solution depending on the
variable order. In our approach we decompose the contribution of the input variables directly by using the correlation information of the standard
normal space. This decomposition results in a unique solution. 
For non-normal distributions the Nataf model \shortcite{Nataf1962} is applied to transform the original space and the correlation 
information to the standard normal space.
The proposed method finally provides modified sampling matrices following the
original joint probability distribution while keeping the samples of a certain variable unchanged. These
samples are used directly to compute the model output without any assumptions or
approximations of the model response function which makes the proposed method model independent.
Additionally, reduced polynomial models are presented which are used to estimate the variance contribution of single variables.

Alternatively to the decomposition approach, in \shortcite{Most_Saltelli_2010} a reordering approach has been proposed.
In this method the investigated variable has been taken as the first one in the variables set for the Cholesky decomposition in the standard normal space.
With this approach the samples required for the estimation of the sensitivity indices can be directly determined. 
This method has been adapted in \shortcite{Kucherenko2012}. In our paper we do not discuss 
this approach, since the implementation is much more complicated compared to the decomposition approach. 

\section{Variance based sensitivity analysis}

\subsection{First order and total effect sensitivity indices}
Assuming a model with a scalar output $Y$ as a function of a given set of $m$ random input parameters $X_i$
\begin{equation}
Y=f(X_1,X_2,\ldots,X_m),
\label{model}
\end{equation}
the first order sensitivity measure was introduced as \shortcite{Sobol1993}
\begin{equation}
S_i=\frac{V_{X_i}(E_{\mathbf{X}_{\sim i}}(Y|X_i))}{V(Y)},
\label{first}
\end{equation}
where $V(Y)$ is the unconditional variance of the model output and 
$V_{X_i}(E_{\mathbf{X}_{\sim i}}(Y|X_i))$ is named the {\it variance of  conditional expectation}
with $\mathbf{X}_{\sim i}$ denoting the matrix of all factors but $X_i$.
$V_{X_i}(E_{\mathbf{X}_{\sim i}}(Y|X_i))$ measures the first order effect of $X_i$ on the model output.

Since first order sensitivity indices measure only the decoupled influence of each variable
an extension for higher order coupling terms is necessary.
For this purpose total effect sensitivity indices have been introduced \shortcite{Homma1996}
\begin{equation}
S_{Ti}=1-\frac{V_{\mathbf{X}_{\sim i}}(E_{X_i}(Y|\mathbf{X}_{\sim i}))}{V(Y)},
\label{total}
\end{equation}
where $V_{\mathbf{X}_{\sim i}}(E_{X_i}(Y|\mathbf{X}_{\sim i}))$ measures the first order effect of $\mathbf{X}_{\sim i}$
on the model output which does not contain any effect corresponding to $X_i$.

\subsection{Sampling based estimates}
\label{sampling}
\begin{figure}[th]
\begin{center}
{\footnotesize
\input{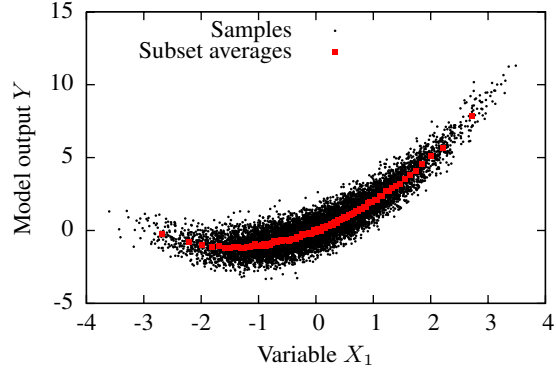}
}
\vspace{-5mm}
\caption{Scatter plot of a single input variable with the model output including subset averages of sorted sample subsets}
\label{anthill}
\end{center}
\end{figure}

One of the most simple procedures to estimate the first order indices
is to sample a set of $n$ samples of the input parameter set $\mathbf{X}= [ X_1, X_2, \ldots, X_m ]$
according to their joint probability distribution and compute for each sample $\mathbf{x}_j$
the model output $y_j$. A scatter plot is obtained if the model output values 
are plotted against the values of a single variable as shown in Figure \ref{anthill}.
Then the sample values are sorted according to a single input variable $X_i$
and subdivided in a number of subsets. For each subset the average value is computed which is equivalent
to $E_{\mathbf{X}_{\sim i}}(Y|X_i)$. By calculating the variance of the subset averages,
an estimate of $V_{X_i}(E_{\mathbf{X}_{\sim i}}(Y|X_i))$ is obtained, which can directly be
used to estimate the first order sensitivity index.
This procedure has the advantage that it gives suitable estimates for independent and dependent input parameters.
However, an extension for the total sensitivity indices does not exist so far since a sorting
concerning $\mathbf{X}_{\sim i}$ seems not possible.

\begin{figure}[th]
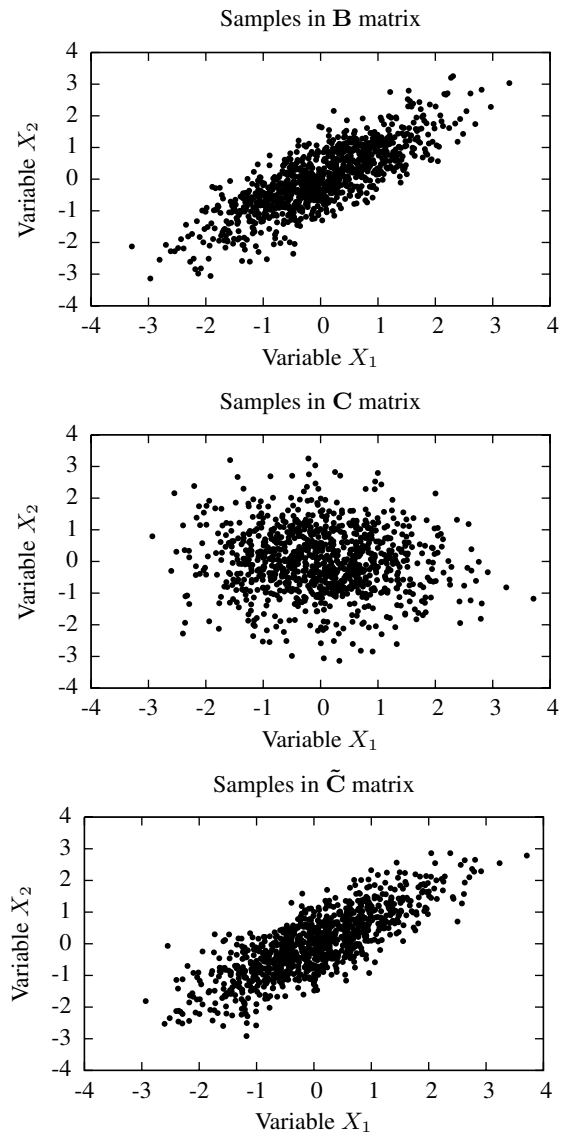

\begin{center}
\vspace{-5mm}
{\footnotesize
\input{pics/correlation_input_B_matrix_mod}\\

\vspace{-3mm}
\input{pics/correlation_input_C_matrix_mod}\\

\vspace{-3mm}
\input{pics/correlation_input_C_matrix_corr_mod}
}
\vspace{-5mm}
\caption{Original and modified joint probability distributions of two correlated variables using the original and the extended matrix combination approach}
\label{anthill_matrix}
\end{center}
\end{figure}

In order to compute the first order and total sensitivity indices using sampling methods,
a matrix combination approach is very common in sensitivity analysis.
In this procedure two independent sampling matrices $\mathbf{A}$ and $\mathbf{B}$
are generated according to the joint probability density function of the input parameters
and recombined in a matrix $\mathbf{C}_i$, where $\mathbf{C}_i$ contains the entries of matrix $\mathbf{B}$
except the $i$-th column which is taken from $\mathbf{A}$.
The estimates of the sensitivity indices can be obtained following \shortcite{Saltelli_2008_Book}
as
\begin{equation}
\hat S_i = \frac{\mathbf{y_A}^T \mathbf{y}_{\mathbf{C}_i} - n(\bar y_\mathbf{A})^2}
{\mathbf{y_A}^T \mathbf{y_A} - n(\bar y_\mathbf{A})^2},\quad
\hat S_{T_i} = 1 - \frac{\mathbf{y_B}^T \mathbf{y}_{\mathbf{C}_i} - n(\bar y_\mathbf{B})^2}
{\mathbf{y_B}^T \mathbf{y_B} - n(\bar y_\mathbf{B})^2},
\label{matrix_estimate}
\end{equation}
where $\mathbf{y_A}$, $\mathbf{y_B}$ and $\mathbf{y}_{\mathbf{C}_i}$ are vectors containing the model outputs
of the sampling matrices and $\bar y_\mathbf{A}$ and $\bar y_\mathbf{B}$ are the corresponding mean value estimates.
Instead of the estimators in Eq.~(\ref{matrix_estimate}) other approaches exist as discussed
in \shortcite{Saltelli2010}. However all of these methods are based on the generation of the sampling matrices $\mathbf{A}$, $\mathbf{B}$ and $\mathbf{C}_i$.
Due to the combination of independent columns of  $\mathbf{A}$ and $\mathbf{B}$ the matrices $\mathbf{C}_i$
follow the joint probability density function of $\mathbf{A}$ and $\mathbf{B}$ only if the input variables are independent.
In case of dependent parameters $X_k$ and $X_l$ the $k$-th and $l$-th columns of the matrices $\mathbf{C}_k$ and $\mathbf{C}_l$
are independent due to the independence of $\mathbf{A}$ and $\mathbf{B}$. Correlations between $X_i$
and the remaining variables are lost in $\mathbf{C}_i$, see the first two plots in Figure \ref{anthill_matrix}.
For the correlated case only a grouping of the dependent parameters according to \shortcite{Jaques2006}
would lead to equivalent probability density functions, but this would limit the method to
special cases where the grouping is possible.


\subsection{Regression based estimates}
\label{regression}

In order to decrease the numerical effort of a probabilistic analysis
often surrogate models are used to approximate the model output instead of evaluating the
real sophisticated model. 
For this purpose, the model output is replaced mainly by a continuous function, which can be evaluated quite fast
compared to a real model call.
A very common method is polynomial regression,
where the model response $Y$ is generally approximated by a polynomial basis function
\begin{equation}
\mathbf{p}_\mathbf{X}^T(\mathbf{x})=\left[ 1\;x_1\;x_2\;x_3\; \ldots \;x_1^2\;x_2^2\;x_3^2\; \ldots\;x_1x_2\;x_1x_3\;\ldots\; x_2x_3\;\ldots\right]
\end{equation}
of linear or quadratic order with or without linear coupling terms.
The model output $y_j$ for a given sample $\mathbf{x}_j$ of the input parameters $\mathbf{X}$
can be formulated as the sum of the approximated value $\hat y_j$ and an error term $\epsilon_j$
\begin{equation}
y(\mathbf{x}_j) = \hat y_j(\mathbf{x}_j) + \epsilon_j = \mathbf{p}_\mathbf{X}^T(\mathbf{x}_j){\boldsymbol\beta}_Y + \epsilon_j,
\end{equation}
where $\boldsymbol{\beta}_Y$ is a vector containing the unknown regression coefficients.
These coefficients are generally estimated from a given set of
sampled support points by
assuming independent errors with equal variance at each point.
By using a matrix notation the resulting least squares solution reads \shortcite{Myers2002}
\begin{equation}
\boldsymbol{\hat \beta}_Y=(\mathbf{P}_\mathbf{X}^T\mathbf{P}_\mathbf{X})^{-1}\mathbf{P}_\mathbf{X}^T \mathbf{y},
\end{equation}
where $\mathbf{P}_\mathbf{X}$ is a matrix containing the basis polynomials of the support point samples.

In order to verify the approximation model the 
coefficient of determination is often used
\begin{equation}
R^2 = 1 - \frac{\sum_{j=1}^n (y_j - \hat y_j)^2}{\sum_{j=1}^n(y_j - \bar y)^2}, \quad \bar y = \frac{1}{n}\sum_{j=1}^n y_j.
\end{equation}
Generally, $R^2$ is interpreted as the fraction of the variance of the true model
represented by the approximation model. This can be used to estimate
total sensitivity indices based on the multivariate regression model \shortcite{Bucher2009_Book}
\begin{equation}
\hat S^R_{T_i} = R^2_{\mathbf{X}} - R^2_{\mathbf{X}_{\sim i}},
\label{R2_sens}
\end{equation}
where $R^2_{\mathbf{X}}$ is obtained using the full parameter set to build up the regression model
and $R^2_{\mathbf{X}_{\sim i}}$ is estimated for a regression model with a reduced parameter set $\mathbf{X}_{\sim i}$.

First order sensitivity indices can be estimated by using only these polynomial basis terms which belong to the investigated variable $X_i$.
The resulting one-dimensional coefficient of determination is a direct estimate of first order indices
\begin{equation}
\hat S^R_{i} = R^2_{X_i}.
\end{equation}

\section{Correlated input variables}
\subsection{Representation of non-normal joint distribution functions}
\label{nataf_section}
In probabilistic models the dependency between input variables is often
modeled by a linear correlation model based on the Pearson correlation coefficient
representing pairwise linear correlations
\begin{equation}
\rho(X_i,X_j) = \frac{E[(X_i-\bar X_i)(X_j-\bar X_j)]}{\sqrt{V(X_i)V(X_j)}},
\end{equation}
with the mean values $\bar X_i$ and $\bar X_j$.

In our study the Nataf model \shortcite{Nataf1962} is used to generate multivariate joint distribution function
of non-normally distributed random variables.
In this model the marginal distributions of the individual random variables are transformed to standard normal densities 
\begin{equation}
Z_i = \mathcal{T}_{X_i}(X_i),
\end{equation} 
and
the resulting multivariate distribution is assumed to be jointly normal
\begin{equation}
f_\mathbf{Z}(\mathbf{z}) = \frac{1}{\sqrt{(2\pi)^m |\mathbf{C}_\mathbf{ZZ}|}}
\exp\left[-\frac{1}{2}\mathbf{z}^T\mathbf{C}_\mathbf{ZZ}^{-1}\mathbf{z}\right],
\label{normal_density}
\end{equation} 
where $m$ is the number and 
$\mathbf{C}_\mathbf{ZZ}$ is the covariance matrix of the random variables which contains in case of standard normal variables 
the correlation coefficients $C_{ij} = \rho(Z_i,Z_j)$.
If the correlation coefficients are given in the standard normal space, the correlation coefficients in the original space can be obtained as follows
\begin{equation}
\rho(X_i,X_j) = \int_{-\infty}^{\infty}\int_{-\infty}^{\infty} \frac{x_i-\bar X_i}{\sigma_{X_i}}\frac{x_j-\bar X_j}{\sigma_{X_j}} f_{Z_iZ_j}(z_i,z_j)dz_idz_j,
\label{nataf}
\end{equation} 
where $f_{Z_iZ_j}(z_i,z_j)$ is a two-dimensional standard normal probability density function.
Often the correlation coefficients are given in the original space. In this case the correlation coefficients in the standard normal space can be derived by solving
Eq.~(\ref{nataf}) iteratively.

\subsection{Decomposition of input variables}
\label{correlation}
If input variables are correlated with each other, a single variable can be
represented as the sum of a function of the remaining inputs (correlated) and an independent (uncorrelated) part
\begin{equation}
X_i=f(X_1,X_2,\ldots,X_{i-1},X_{i+1},\ldots,X_m) + X_i^{U,{\mathbf{X}_{\sim i}}}
=X_i^{C,{\mathbf{X}_{\sim i}}} + X_i^{U,{\mathbf{X}_{\sim i}}}.\\
\label{correlated_input}
\end{equation}
By using a linear correlation model,
the correlated part of each input  variable can be represented 
by a linear combination of the remaining variable set ${\mathbf{X}_{\sim i}}$
\begin{equation}
X_i^{C,{\mathbf{X}_{\sim i}}} = \beta_{X_i,0} + \sum_{j=1,j\neq i}^{m} \beta_{X_i,j} X_j.
\end{equation}
In case of standard normal and linearly correlated random variables $Z_i$,
the coefficients 
\begin{equation}
Z_i^{C,{\mathbf{Z}_{\sim i}}} = \beta_{Z_i,0} + \sum_{j=1,j\neq i}^{m} \beta_{Z_i,j} Z_j, \quad 
\end{equation}
can be directly determined from the correlation coefficients
\begin{equation}
\beta_{Z_i,0} = 0, \quad \boldsymbol{\beta}_{Z_i} = \mathbf{C}_{\mathbf{Z}_{\sim i}\mathbf{Z}_{\sim i}}^{-1} \boldsymbol{\rho}_{Z_i},
\label{decomposition_corr}
\end{equation}
where $\boldsymbol{\beta}_{Z_i}= [ \beta_{Z_i,1}, \ldots, \beta_{Z_i,i-1}, \beta_{Z_i,i+1}, \ldots, \beta_{Z_i,m} ]$ contains the coefficients belonging to 
$\mathbf{Z}_{\sim i}$, $\mathbf{C}_{\mathbf{Z}_{\sim i}\mathbf{Z}_{\sim i}}$ is the correlations matrix of $\mathbf{Z}_{\sim i}$ 
and $\boldsymbol{\rho}_{Z_i} = [ \rho(Z_i,Z_1), \ldots, \rho(Z_i,Z_{i-1}), \rho(Z_i,Z_{i+1}), \ldots, \rho(Z_i,Z_m) ]$ contains the correlation coefficients of $\mathbf{Z}_{\sim i}$
with respect to $Z_i$.
A derivation of Eq.~(\ref{decomposition_corr}), which is valid only in the standard normal space, can be found in the appendix.

For a given set of discrete samples of the standard normal parameter set $\mathbf{Z}$ arranged in matrix $\mathbf{M}$,
the uncorrelated parts of the $j$th column of $\mathbf{M}$ with respect $\mathbf{Z}_{\sim i}$ can be determined as follows
\begin{equation}
\mathbf{M}_{(j)}^{U,{\mathbf{Z}_{\sim i}}} = \mathbf{M}_{(j)} - \mathbf{M}_{(j)}^{C,{\mathbf{Z}_{\sim i}}}
\label{uncorrelated_M}
\end{equation}
with
\begin{equation}
\mathbf{M}_{(j=i)}^{C,{\mathbf{Z}_{\sim i}}} = \sum_{k=1,k\neq i}^{m}\beta_{Z_i,k}\mathbf{M}_{(k)}, 
\quad \mathbf{M}_{(j\neq i)}^{C,{\mathbf{Z}_{\sim i}}}= \mathbf{M}_{(j)}.
\end{equation}

For later use the correlated and uncorrelated parts of a single random variable with respect to only one other
variable is introduced.
The correlated part of $X_j$ with respect to $X_i$ 
can be formulated as follows
\begin{equation}
X_j^{C, X_i} = \beta_{X_j,0}^{X_i} + \beta_{X_j,1}^{X_i} X_i.
\label{1d_regression}
\end{equation}
In the standard normal space we obtain
\begin{equation}
Z_j^{C, Z_i} = \beta_{Z_j,0}^{Z_i} + \beta_{Z_j,1}^{Z_i} Z_i, \quad \beta_{Z_j,0}^{Z_i} = 0, \quad \beta_{Z_j,i}^{Z_i} = \rho(Z_i,Z_j).
\end{equation}
For a given sampling matrix $\mathbf{M}$ the decomposition reads
\begin{equation}
\mathbf{M}_{(j)}^{U,Z_i} = \mathbf{M}_{(j)} - \mathbf{M}_{(j)}^{C,Z_i}
= \mathbf{M}_{(j)} - \rho(Z_i,Z_j)\mathbf{M}_{(i)}.
\label{uncorrelated_M_1D}
\end{equation}
In the second decomposition the corresponding $i$th column of $\mathbf{M}^{C, Z_i}$ is equal to $\mathbf{M}_{(i)}$ and the values of the
$i$th column of $\mathbf{M}^{U, X_i}$ vanish.

As discussed in \shortcite{Saltelli2004} the conditional variance $V_{\mathbf{X}_{\sim i}}(E_{X_i}(Y|\mathbf{X}_{\sim i}))$
may decrease by introducing dependence between $X_i$ and $\mathbf{X}_{\sim i}$. This may result in total effect indices
smaller than first order indices. Furthermore, if we use a linear correlation model and increase the absolute value of the correlation 
coefficient between two input variables close to one, the total effect indices of both variables approach to zero.
For factor fixing this relation requires a step-wise fixing of a single factor and a revaluation of the indices of the remaining variables.
If only the sensitivity indices of the full model are used for factor fixing, this may result in a mistaken removing of important variables.
One possibility to overcome this, is to evaluate the sensitivity indices on one side for the uncorrelated part of a single variable $X_i^{U}$
and on the other side for this variable including all correlated portions of the other variables $X_j^{C, X_i}, j\neq i$.
This concept was introduced for first order indices of linear models in \shortcite{Xu2008}.
In our study we introduce first order indices $S_i^C$ and $S_i^U$ as well as total effect indices $S_{T_i}^C$ and $S_{T_i}^U$
for the correlated and uncorrelated parts of a single variable $X_i$.
Per definition $S_i^C$ is equivalent to the original first order index and $S_{T_i}^U$ to the original total effect index.
The additional measures $S_i^U$ and especially $S_{T_i}^C$ should give us additional information for factor fixing without 
requiring a step-wise revaluation of the sensitivity indices.


\section{Estimation of sensitivity indices for correlated input variables}
\subsection{Extension of the matrix combination approach}
\label{extension_sobol}
In the original matrix combination approach presented in section \ref{sampling} it is intended to
modify a single input variable $X_i$ by a second random sampling set
while keeping the other variables
unchanged in order to add or to remove the influence of $X_i$ from the model.

In order to calculate the correlated first order and total effect indices, not only the samples of $X_i$ itself, but also all correlated
parts of the remaining variables with respect to $X_i$ have to be modified.
By assuming a linear correlation model in the standard normal space as presented in section \ref{correlation},
a decomposition of the matrices $\mathbf{A}$ and $\mathbf{B}$ in an uncorrelated and correlated part with respect to $X_i$
is performed. For this purpose both matrices are transformed to the correlated standard normal space by the marginal transformations
\begin{equation}
\mathcal{A}_{kl} = \Phi^{-1}[F_{X_l}(A_{kl})],\quad \mathcal{B}_{kl} = \Phi^{-1}[F_{X_l}(B_{kl})],
\end{equation} 
where $\Phi^{-1}(\cdot)$ is the inverse  cumulative distribution function of a standard normal random variable and $F_{X_l}(\cdot)$ is
the cumulative distribution function of $X_l$.
By using the decomposition proposed in Eq.~(\ref{uncorrelated_M_1D}) we obtain the columns of the correlated and uncorrelated sampling matrices as follows
\begin{eqnarray}
\boldsymbol{\mathcal{A}}_{(j)}^{C, Z_i} &=& \rho(Z_i,Z_j)\boldsymbol{\mathcal{A}}_{(i)},\quad
\boldsymbol{\mathcal{A}}_{(j)}^{U, Z_i} = \boldsymbol{\mathcal{A}}_{(j)} - \boldsymbol{\mathcal{A}}_{(j)}^{C, Z_i},\\
\boldsymbol{\mathcal{B}}_{(j)}^{C, Z_i} &=& \rho(Z_i,Z_j)\boldsymbol{\mathcal{B}}_{(i)},\quad
\boldsymbol{\mathcal{B}}_{(j)}^{U, Z_i} = \boldsymbol{\mathcal{B}}_{(j)} - \boldsymbol{\mathcal{B}}_{(j)}^{C, Z_i}.
\end{eqnarray} 
Now a modified matrix $\boldsymbol{\tilde \mathcal{C}}^{C}_i$ can be obtained, which contains the uncorrelated part of $\boldsymbol{\mathcal{B}}$ 
with respect to $Z_i$
and the correlated part of $\boldsymbol{\mathcal{A}}$
\begin{equation}
\boldsymbol{\tilde \mathcal{C}}_{i}^{C} = \boldsymbol{\mathcal{C}}_{i}^{C, Z_i} = \boldsymbol{\mathcal{B}}^{U, Z_i} + \boldsymbol{\mathcal{A}}^{C, Z_i}.
\label{decomposition_C}
\end{equation}
Finally the entries of matrix $\boldsymbol{\tilde \mathcal{C}}^{C}_i$ are transformed to the original space by the individual marginal transformations
\begin{equation}
\tilde C_{i,kl}^{C} = F_{X_l}^{-1}[\Phi(\tilde \mathcal{C}_{i,kl}^{C})].
\end{equation} 
Since the correlated part with respect to $X_i$
is modified simultaneously for all variables, $\mathbf{\tilde C}_i^C$ follows the original joint probability distribution of $\mathbf{X}$
as shown in Figure \ref{anthill_matrix}.

In order to obtain the uncorrelated first order and total effect indices we use the decomposition with respect to $\mathbf{X}_{\sim i}$.
For this purpose
the matrices $\boldsymbol{\mathcal{A}}$ and $\boldsymbol{\mathcal{B}}$ are decomposed following Eq.~(\ref{uncorrelated_M})
\begin{eqnarray}
\boldsymbol{\mathcal{A}}_{(j)}^{U, {\mathbf{Z}_{\sim i}}} = \boldsymbol{\mathcal{A}}_{(j)} - \boldsymbol{\mathcal{A}}_{(j)}^{C, {\mathbf{Z}_{\sim i}}},\quad
\boldsymbol{\mathcal{A}}_{(j=i)}^{C, {\mathbf{Z}_{\sim i}}} &=& \sum_{k=1,k\neq i}^{m}\beta_{Z_i,k}\boldsymbol{\mathcal{A}}_{(k)},\quad 
\boldsymbol{\mathcal{A}}_{(j\neq i)}^{C, {\mathbf{Z}_{\sim i}}} = \boldsymbol{\mathcal{A}}_{(j)},\\
\boldsymbol{\mathcal{B}}_{(j)}^{U, {\mathbf{Z}_{\sim i}}} = \boldsymbol{\mathcal{B}}_{(j)} - \boldsymbol{\mathcal{B}}_{(j)}^{C, {\mathbf{Z}_{\sim i}}},\quad
\boldsymbol{\mathcal{B}}_{(j=i)}^{C, {\mathbf{Z}_{\sim i}}} &=& \sum_{k=1,k\neq i}^{m}\beta_{Z_i,k}\boldsymbol{\mathcal{B}}_{(k)},\quad
\boldsymbol{\mathcal{B}}_{(j\neq i)}^{C, {\mathbf{Z}_{\sim i}}} = \boldsymbol{\mathcal{B}}_{(j)}.
\end{eqnarray} 
The resulting $\boldsymbol{\tilde \mathcal{C}}^{U}_i$ contains the uncorrelated part of $Z_i$ from matrix $\boldsymbol{\mathcal{A}}$ and the 
correlated part with respect to $\mathbf{Z}_{\sim i}$
from matrix $\boldsymbol{\mathcal{B}}$
\begin{equation}
\boldsymbol{\tilde \mathcal{C}}_{i}^{U} = \boldsymbol{\mathcal{C}}_{i}^{U, \mathbf{Z}_{\sim i}} = 
\boldsymbol{\mathcal{A}}^{U, \mathbf{Z}_{\sim i}} + \boldsymbol{\mathcal{B}}^{C, \mathbf{Z}_{\sim i}}.
\label{decomposition_C2}
\end{equation}
Analogous to the correlated part, $\mathbf{\tilde C}_i^U$ is obtained by transforming the entries of $\boldsymbol{\tilde \mathcal{C}}^{U}_i$ back to the origin
space.

\subsection{Extension for regression based indices}
The regression based indices presented in section \ref{regression}
can be directly applied for correlated input variables.
However, the one-dimensional coefficient of determination is an estimate of the first order index of a single variable including all correlated parts of the other
variables. Using a matrix $\boldsymbol{\mathcal{Z}}$ containing the support points of the regression in the standard normal space, the first order index reads
\begin{equation}
\hat S^{R,C}_{i} = R^2_{\boldsymbol{\mathcal{Z}},Z_i}.
\end{equation}
The estimate for the total effect index given in Eq.~(\ref{R2_sens}) quantifies the total variance contribution of the uncorrelated part $Z_i^U$ of variable $Z_i$
\begin{equation}
\hat S^{R,U}_{T_i} = R^2_{\boldsymbol{\mathcal{Z}},\mathbf{Z}} - R^2_{\boldsymbol{\mathcal{Z}},\mathbf{Z}_{\sim i}}.
\end{equation}
The correlated part of $Z_i$ can be represented by the other variables and thus the explained variation of the reduced model 
is decreased by the contribution of $Z_i^U$ only.
The total effect of the variable $Z_i$ including all correlated parts of the other variables can be estimated analogously to the matrix combination approach:
The sampling matrix $\boldsymbol{\mathcal{Z}}$ is decomposed in a correlated and uncorrelated part with respect to $Z_i$ according to Eq.~(\ref{decomposition_C2}).
Then the uncorrelated part of the samples $\boldsymbol{\mathcal{Z}}^{U, Z_i}$ is used for the reduced model within the estimate of the total effect index 
\begin{equation}
\hat S^{R,C}_{T_i} = R^2_{\boldsymbol{\mathcal{Z}},\mathbf{Z}} - R^2_{\boldsymbol{\mathcal{Z}}^{U, Z_i},\mathbf{Z}_{\sim i}}.
\end{equation}

In order to estimate the first order index of the uncorrelated part of the variable $Z_i$ with respect to all other variables,
Eq.~(\ref{uncorrelated_M}) is used to calculate the uncorrelated part of $i$th column of the sample matrix $\boldsymbol{\mathcal{Z}}$
and the corresponding first order sensitivity index is obtained
\begin{equation}
\hat S^{R,U}_{i} = R^2_{\boldsymbol{\mathcal{Z}}^{U, \mathbf{Z}_{\sim i}},Z_i}.
\end{equation}

With the four measures $\hat S^{R,U}_{i}$, $\hat S^{R,C}_{i}$, $\hat S^{R,U}_{T_i}$ and $\hat S^{R,C}_{T_i}$
regression based sensitivity indices are introduced, which give similar results as the model independent measures proposed in section \ref{extension_sobol}, but only
if the regression model can represent the underlying investigated model.
If this is not the case, e.g. if complex nonlinearities and interactions describe the model response $Y$ in terms of the input variables $\mathbf{X}$,
the polynomial based measures quantify only the contribution which is represented by the basis function.
The total contribution  of higher order and unexplained dependencies, which are not represented by the regression model, can be estimated using the full regression basis
\begin{equation}
\hat S^{Unexplained}_{T_i} = 1-R^2_{\boldsymbol{\mathcal{Z}},\mathbf{Z}}.
\end{equation}

Since the introduced sensitivity measures investigate the influence of a single variable including all correlated parts of the other variables,
it can not be distinguished, if the influence of this variable is caused by its contribution within the model or if it is caused by its correlation 
with other important variables.
This fact shall be clarified by a  simple example:
A purely additive model with three inputs is given
\begin{equation}
Y =\beta_0  + \beta_1 X_1+\beta_2 X_2+ \beta_3 X_3.
\end{equation}
The input variables are normally distributed, have zero mean and the covariance
\begin{equation}
\Gamma = \left[
\begin{array}{ccc}
\sigma_{X_1}^2 & 0 & 0\\
0 & \sigma_{X_2}^2 & \rho \sigma_{X_2}\sigma_{X_3}\\
0 & \rho \sigma_{X_2}\sigma_{X_3} & \sigma_{X_3}^2\\
 \end{array}\right],
\end{equation}
where $\rho$ is the linear correlation between $X_2$ and $X_3$.

The first order and total effect indices of this problem can be derived analytically as follows:
$X_3$ can be formulated with respect to $X_2$ and vice versa as 
\begin{eqnarray}
X_3 &= \rho \frac{\sigma_{X_3}}{\sigma_{X_2}} X_2 + \sqrt{1-\rho^2} X_3^U, \nonumber\\
X_2 &= \rho \frac{\sigma_{X_2}}{\sigma_{X_3}} X_3 + \sqrt{1-\rho^2} X_2^U, 
\label{dependent}
\end{eqnarray}
with $X_2^U$ and $X_3^U$ having the same variances as $X_2$ and $X_3$, respectively.
Since $X_2$ and $X_3^U$ as well as  $X_3$ and $X_2^U$ are independent, we can calculate
the variance contribution of the inputs as follows
\begin{eqnarray}
V_{X_1^C}(Y) &=& V_{X_1^U}(Y) = \beta_1^2 \sigma_{X_1}^2,\nonumber\\
V_{X_2^C}(Y) &=& (\beta_2\sigma_{X_2}+\rho\beta_3\sigma_{X_3})^2,\nonumber\\
V_{X_2^U}(Y) &=& (1-\rho^2)\beta_2^2\sigma_{X_2}^2,\nonumber\\
V_{X_3^C}(Y) &=& (\beta_3\sigma_{X_3}+\rho\beta_2\sigma_{X_2})^2,\nonumber\\
V_{X_3^U}(Y) &=& (1-\rho^2)\beta_3^2\sigma_{X_3}^2.\label{variance_contribution}
\end{eqnarray}
With help of the total variance
\begin{equation}
V(Y)= V_{X_1}(Y)+V_{X_2^C}(Y)+V_{X_3^U}(Y)=V_{X_1}(Y)+V_{X_2^U}(Y)+V_{X_3^C}(Y),
\end{equation}
the sensitivity indices can be determined
\begin{equation}
S^C_i = S_{T_i}^C = \frac{V_{X_i}^C(Y)}{V(Y)}, \quad S^U_i = S_{T_i}^U = \frac{V_{X_i}^U(Y)}{V(Y)}.
\end{equation}
From Eq.~(\ref{variance_contribution}) it follows, that the variance contribution of the uncorrelated parts of $X_2$ and $X_3$
vanish if the correlation coefficient tends to one or minus one.
On the other hand it can be seen, that if one of the variable factors $\beta_2$ or $\beta_3$ is zero,
the corresponding variance contribution and thus the sensitivity index of the correlated part is not zero,
due to the remaining influence of the other variable.

In order to distinguish between the variance contribution due to the correlation and the contribution by the model parameters,
an effective variance contribution is defined
\begin{equation}
V_{X_i}^{eff}(Y) = \beta_i^2\sigma_{X_i}^2.
\end{equation}

For the general polynomial regression case, the following procedure is proposed: 
First the full regression model is build up in the standard normal space and the regression coefficients $\hat {\boldsymbol \beta}$ are determined.
Then the first order effects are estimated by using only the linear and quadratic coefficients of the full model, which belong to the 
investigated variable $Z_i$,
\begin{equation}
\hat S^{\boldsymbol \beta}_{i} = \frac{V_{{\boldsymbol \beta}_{Z_i}}(Y)}{V(Y)},
\end{equation}
where $V_{{\boldsymbol \beta}_{Z_i}}(Y)$ is the resulting variance if only the first order regression terms of $Z_i$ are considered.
Total effects can be estimated similarly
\begin{equation}
\hat S^{\boldsymbol \beta}_{T_i} = 1-\frac{V_{{\boldsymbol \beta}_{\mathbf{Z}_{\sim i}}}(Y)}{V(Y)},
\end{equation}
where $V_{{\boldsymbol \beta}_{\mathbf{Z}_{\sim i}}}(Y)$ is estimated by using the regression model without all coefficients, first order and interaction,
which belong to ${Z_i}$.
The definition of $S^{\boldsymbol \beta}_{i}$ and $S^{\boldsymbol \beta}_{T_i}$ is not following the general definition of the first order and total effect
sensitivity indices, where the variance of conditional expectation is used. Nevertheless, this measures may be used to 
give additional information about the unknown underlying model.

However, for strongly correlated random variables the accuracy of the estimated regression coefficients may be very poor.
In such a case the estimated sensitivity measures $\hat S^{\boldsymbol \beta}_{i}$ and $\hat S^{\boldsymbol \beta}_{T_i}$ may have very low accuracy.
In the first numerical example, this problem is investigated.

\section{Numerical examples}
\subsection{Additive linear model}
In the first example a purely additive model is investigated
\begin{equation}
Y = X_1+X_2+X_3.
\end{equation}
The input variables are normally distributed, have zero mean and the covariance
\begin{equation}
\Gamma = \left[
\begin{array}{ccc}
1 & 0 & 0\\
0 & 1 & 2\rho \\
0 & 2\rho & 4\\
 \end{array}\right].
\end{equation}
Different cases with  $\rho = 0.0, 0.8, 0.99999, -0.8, -0.5$ are investigated with the sampling based approach and the regression based approach.
The sampling based sensitivity indices are calculated using the subset averaging and the
proposed matrix recombination method with 10.000 Latin Hypercube samples.
In Table~\ref{sensitivity_additive} the estimated sensitivity indices are compared to the analytical solution obtained with Eq.~(\ref{variance_contribution}).
\begin{table}[th]
\tiny
\begin{tabular}{lccccccccccccccccc}
    \hline
	& \multicolumn{2}{c}{Analytical} &&  Subset && \multicolumn{4}{c}{Matrix combination} && \multicolumn{6}{c}{Regression}\\
	 & $S_i^C, S_{T_i}^C$ & $S_i^U, S_{T_i}^U$  &&$\hat S_i$  && $\tilde S_i^C$ & $\tilde S_{T_i}^C$ & $\tilde S_i^U$ & $\tilde S_{T_i}^U$ 
	 && $\hat S_i^{R,C}$ & $\hat S_{T_i}^{R,C}$ & $\hat S_i^{R,U}$ & $\hat S_{T_i}^{R,U}$  & $\hat S_i^{\boldsymbol \beta}$ & $\hat S_{T_i}^{\boldsymbol \beta}$ \\
   \hline
	$\rho = 0.0$ \rule[0mm]{0pt}{2.5ex}\\
   $X_1$ & 0.167& 0.167 && 0.177 && 0.157 & 0.168 & 0.157 & 0.166 && 0.155 & 0.155 & 0.168 & 0.168 & 0.168 & 0.155 \\
   $X_2$ & 0.167& 0.167 && 0.177 && 0.154 & 0.164 & 0.153 & 0.163 && 0.168 & 0.168 & 0.169 & 0.168 & 0.168 & 0.168 \\
   $X_3$ & 0.667& 0.667 && 0.674 && 0.667 & 0.673 & 0.669 & 0.672 && 0.670 & 0.670 & 0.672 & 0.672 & 0.672 & 0.670 \\
   \hline																					 	     	     	 
	$\rho = 0.8$ \rule[0mm]{0pt}{2.5ex}\\													 	     	     	 
   $X_1$ & 0.109& 0.109 && 0.118 && 0.117 & 0.108 & 0.118 & 0.107 && 0.110 & 0.110 & 0.109 & 0.108 & 0.109 & 0.110  \\
   $X_2$ & 0.735& 0.039 && 0.742 && 0.738 & 0.727 & 0.050 & 0.039 && 0.735 & 0.735 & 0.040 & 0.039 & 0.109 & 0.456  \\
   $X_3$ & 0.852& 0.157 && 0.861 && 0.858 & 0.848 & 0.169 & 0.157 && 0.853 & 0.853 & 0.156 & 0.156 & 0.434 & 0.783  \\
   \hline																							  		 	 
	$\rho = 0.99999$ \rule[0mm]{0pt}{2.5ex}\\												         	     	 
   $X_1$ & 0.100& 0.100	&& 0.110 && 0.088 & 0.103 & 0.088 & 0.101 && 0.099 & 0.099 & 0.101 & 0.100 & 0.100 & 0.099 \\
   $X_2$ & 0.900& 0.000	&& 0.907 && 0.900 & 0.916 & 0.000 & 0.000 && 0.900 & 0.900 & 0.001 & 0.000 & 0.100 & 0.500 \\
   $X_3$ & 0.900& 0.000	&& 0.907 && 0.900 & 0.916 & 0.000 & 0.000 && 0.900 & 0.900 & 0.001 & 0.000 & 0.401 & 0.800 \\
   \hline																							  
 	$\rho = -0.8$ \rule[0mm]{0pt}{2.5ex}\\
   $X_1$ & 0.357& 0.357 && 0.368 && 0.361 & 0.359 & 0.359 & 0.359 && 0.362 & 0.362 & 0.354 & 0.353 & 0.353 & 0.362 \\
   $X_2$ & 0.129& 0.129 && 0.137 && 0.141 & 0.130 & 0.136 & 0.139 && 0.134 & 0.134 & 0.127 & 0.127 & 0.353 & -0.788\\
   $X_3$ & 0.514& 0.514 && 0.523 && 0.527 & 0.513 & 0.520 & 0.512 && 0.521 & 0.521 & 0.509 & 0.509 & 1.420 & 0.300 \\
   \hline
 	$\rho = -0.5$ \rule[0mm]{0pt}{2.5ex}\\
   $X_1$ & 0.250& 0.250 && 0.256 && 0.240 & 0.250 & 0.243 & 0.249 && 0.241 & 0.241 & 0.254 & 0.253 & 0.253 & 0.241 \\
   $X_2$ & 0.000& 0.188 && 0.011 && 0.005 & 0.000 & 0.179 & 0.188 && 0.000 & 0.000 & 0.189 & 0.188 & 0.253 & -0.266\\
   $X_3$ & 0.563& 0.750 && 0.573 && 0.571 & 0.570 & 0.755 & 0.752 && 0.564 & 0.564 & 0.759 & 0.759 & 1.023 & 0.495 \\
  \hline
\end{tabular}
\caption{Sensitivity indices of the additive model assuming uncorrelated and correlated inputs}
\label{sensitivity_additive}
\end{table}
The table indicates a very good agreement of all estimated indices with the
reference values for the uncorrelated
 and the correlated cases.
In the uncorrelated case, the sensitivity indices of an additive model sum up to one.
By considering correlated inputs this is not the case,
since coupling terms caused by the correlations are introduced indirectly into the model.
This can result in a sum larger than one (correlated formulation with $\rho=0.8$) and also smaller than one
(correlated formulation with $\rho=-0.5$, uncorrelated formulation with $\rho=0.8$). 	
Interestingly the influence of a variable seems to vanish looking only on the correlated or uncorrelated formulation of the total effect indices
which is observed for variable $X_2$ in the correlated formulation with $\rho=-0.5$ and $X_2$ and $X_3$ in the uncorrelated formulation with $\rho=1.0$.
However, if both the correlated and uncorrelated formulation of the total effect indices are considered,
no important variable can be misinterpreted or mistakenly removed from the model. This is not the case if either only the correlated or the uncorrelated
formulation is used.

Additionally the regression based indices by using 1000 Latin Hypercube samples as support points are given in Table \ref{sensitivity_additive}.
The estimates $\hat S^{R,U}_{i}$, $\hat S^{R,C}_{i}$, $\hat S^{R,U}_{T_i}$ and $\hat S^{R,C}_{T_i}$ agree very well with the analytical values.
The introduced measures $\hat S^{\boldsymbol \beta}_{i}$ and $\hat S^{\boldsymbol \beta}_{T_i}$, which analyze the variance contribution by coefficient removing,
are equivalent for the uncorrelated case. In the correlated case, the results differ in that way, that these indices indicate how the total variance
is decreased or eventually increased, if the variable $X_i$ is removed completely from the model formulation.
For this reason negative indices or values larger as one are possible in contrast to classical sensitivity indices.
This measures should be considered as additional information in order to detect imaginary variable importance caused only by input correlations.

\begin{figure}[th]
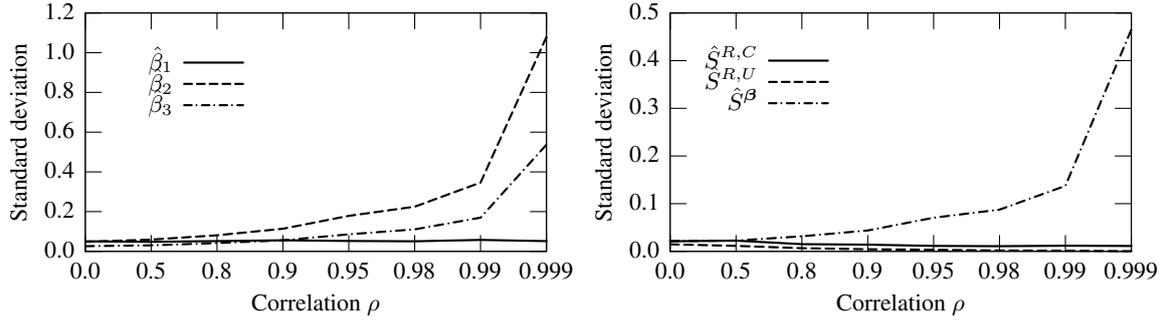

\begin{center}
{\footnotesize
\input{pics/coefficients_accuracy_mod}
\input{pics/indices_accuracy_mod}
}
\vspace{-5mm}
\caption{Standard deviation of the regression coefficients (left) and the estimated total effect sensitivity indices of $X_2$ (right)  
dependent on the correlation for the additive model including random noise}
\label{coefficient_accuracy}
\end{center}
\end{figure}

However, the accuracy of the coefficient based indices $\hat S^{\boldsymbol \beta}_{i}$ and $\hat S^{\boldsymbol \beta}_{T_i}$
may be very low for highly correlated variables, 
if the regression model can not represent the true model perfectly.
This situation is analyzed by adding to the deterministic model a random noise $\epsilon$ which is independent of the inputs
\begin{equation}
Y = X_1+X_2+X_3+\epsilon,
\end{equation}
where $\epsilon$ is a zero-mean normally distributed random variable with standard deviation $\sigma_\epsilon=0.5$.							 
In Figure \ref{coefficient_accuracy} the standard deviation of the estimated regression coefficients of the tree inputs are shown for a set of 100 support point
samples.			   
The figure indicates an increasing error in the regression coefficients of both correlated variables if the correlation coefficient is increased,
while the error in the coefficient of the 
uncorrelated variable is almost unchanged. This results in an inaccurate estimate of the sensitivity measures $\hat S^{\boldsymbol \beta}_{i}$
and $\hat S^{\boldsymbol \beta}_{T_i}$ for highly correlated variables. In Figure \ref{coefficient_accuracy}
the standard deviation of the estimated index $\hat S^{\boldsymbol \beta}_{T_2}$ is displayed additionally. Similar to the regression coefficient itself,
the accuracy of the sensitivity index decreases with increasing correlation. Nevertheless, if the variables are correlated with $\rho \leq 0.9$
the error in the estimate seems acceptable.
In contrast to the measure based on single regression coefficients, the accuracy of the sensitivity indices estimated as the difference of the coefficient of determination
of the full and a reduced regression model is not decreased with increasing correlation. This is plotted additionally in Figure \ref{coefficient_accuracy}.
Using only 100 samples, the accuracy of the estimated regression based measures seems very good.
							   
\subsection{Coupled nonlinear model}
In the second example a nonlinear model is investigated, which contains linear, quadratic and interaction terms
\begin{equation}
Y = X_1+2X_1^2+X_2+X_3+X_2 X_3.
\label{coupled}
\end{equation}
The covariance is taken analogous to the previous example.  							 
\begin{figure}[th]
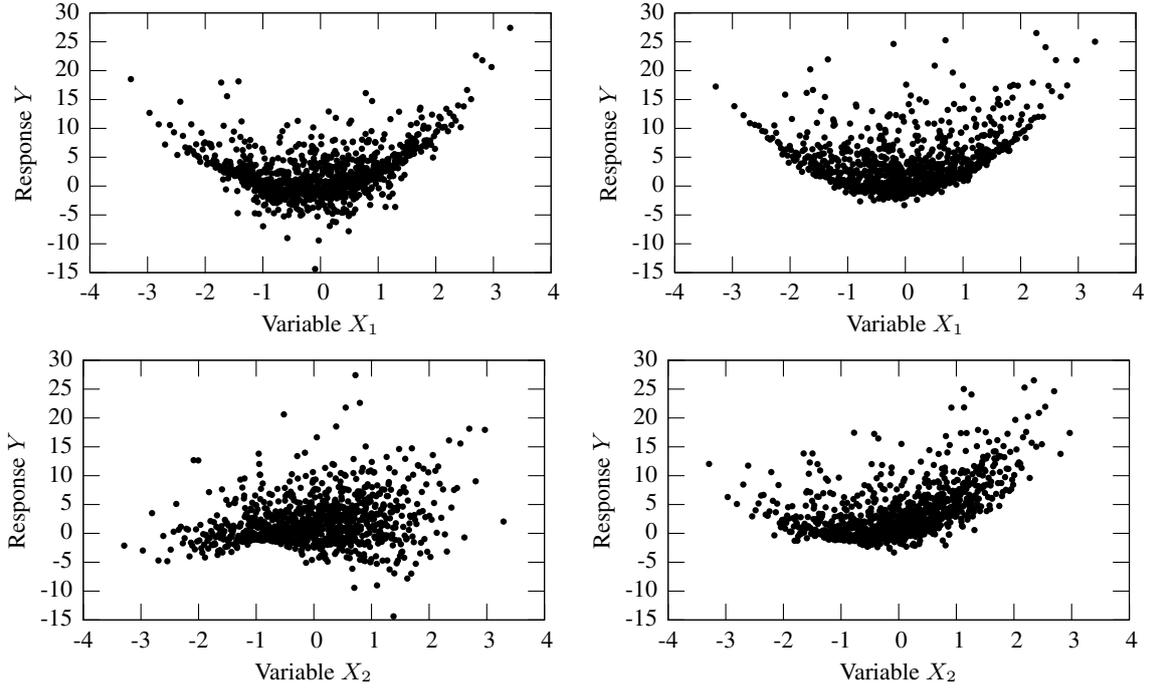

\begin{center}
{\footnotesize
\input{pics/coupled_uncorrelated_X1_mod}
\input{pics/coupled_correlated_X1_mod}\\
\input{pics/coupled_uncorrelated_X2_mod}
\input{pics/coupled_correlated_X2_mod}
}
\vspace{-5mm}
\caption{Anthill plots of the response of the coupled nonlinear function with respect to $X_1$ and $X_2$ by assuming $\rho=0.0$
(left) and $\rho=0.8$ (right)}
\label{anthill_coupled}
\end{center}
\end{figure}
In Figure~\ref{anthill_coupled} the scatter plots are shown for the first two variables.
In the figure a significant quadratic influence of $X_1$ and a significant influence of the coupling term $X_2 X_3$
can be remarked in the uncorrelated case. 
If $X_2$ and $X_3$ are assumed to be correlated, the scatter plot of $X_2$ with $Y$ indicates also a quadratic behavior,
while in the model formulation only a linear and a coupling term appears.
This quadratic influence can be explained, if we rewrite Eq.~(\ref{coupled}) by considering the decomposition of $X_3$ analogous to Eq.~(\ref{dependent})
as follows
\begin{eqnarray}
Y= X_1+2X_1^2+X_2 + \rho \frac{\sigma_{X_3}}{\sigma_{X_2}} X_2 + \sqrt{1-\rho^2} X_3^U + \rho \frac{\sigma_{X_3}}{\sigma_{X_2}}X_2^2+\sqrt{1-\rho^2}X_2X_3^U.
\end{eqnarray}

\begin{table}[th]
\tiny
\begin{tabular}{lccccccccccccccccc}
    \hline
	& Subset && \multicolumn{4}{c}{Matrix combination} && \multicolumn{6}{c}{Regression}\\
	 & $\hat S_i$  && $\tilde S_i^C$ & $\tilde S_{T_i}^C$ & $\tilde S_i^U$ & $\tilde S_{T_i}^U$ 
	 && $\hat S_i^{R,C}$ & $\hat S_{T_i}^{R,C}$ & $\hat S_i^{R,U}$ & $\hat S_{T_i}^{R,U}$  & $\hat S_i^{\boldsymbol \beta}$ & $\hat S_{T_i}^{\boldsymbol \beta}$ \\
   \hline
	$\rho = 0.0$ \rule[0mm]{0pt}{2.5ex}\\
   $X_1$ & 0.484 && 0.489 & 0.518 & 0.490 & 0.518 && 0.491 & 0.511 & 0.491 & 0.510 & 0.512 & 0.491  \\
   $X_2$ & 0.070 && 0.036 & 0.256 & 0.034 & 0.259 && 0.060 & 0.292 & 0.059 & 0.291 & 0.058 & 0.274  \\
   $X_3$ & 0.231 && 0.217 & 0.452 & 0.216 & 0.454 && 0.204 & 0.441 & 0.204 & 0.440 & 0.231 & 0.436  \\
   \hline																					 	     	     	 
	$\rho = 0.8$ \rule[0mm]{0pt}{2.5ex}\\													 	     	     	 
   $X_1$ & 0.382 && 0.405 & 0.362 & 0.405 & 0.361 && 0.373 & 0.367 & 0.371 & 0.365 & 0.365 & 0.373 \\
   $X_2$ & 0.501 && 0.528 & 0.540 & 0.055 & 0.071 && 0.506 & 0.562 & 0.014 & 0.072 & 0.041 & 0.469 \\
   $X_3$ & 0.551 && 0.548 & 0.592 & 0.090 & 0.123 && 0.565 & 0.615 & 0.064 & 0.120 & 0.165 & 0.593 \\
  \hline
\end{tabular}
\caption{Sensitivity indices of the coupled nonlinear model assuming uncorrelated and correlated inputs}
\label{sensitivity_coupled}
\end{table}
In Table~\ref{sensitivity_coupled} the corresponding sensitivity indices are given.
The table indicates a good agreement of the first order indices estimated with the extended matrix combination approach and the regression based method
with the indices obtained with the subset averaging.
The effect seen in the scatter plots, that the influence of $X_2$ dominated in the uncorrelated case by the coupling term
and in the correlated case by quadratic dependencies, can be observed also in the sensitivity indices. In the uncorrelated case
the difference between first order and total effect indices of $X_2$ is more than $20\%$, but in the correlated case this difference decreases significantly.						 
This example clarifies, that the proposed measures quantifying the total effect influence give useful information for model interpretation
  in the case of correlated inputs.

\subsection{Ishigami function}
In the final example the well-known Ishigami function \shortcite{Ishigami1990} is investigated.
This function was defined for independent uniformly distributed variables
\begin{equation}
-\pi \leq X_i \leq \pi, \quad  i=1,2,3,
\end{equation}
as follows
\begin{equation}
Y = \sin(X_1)+a\sin^2(X_2)+bX_3^4\sin(X_1),
\end{equation}
with $a=7.0$ and $b=0.1$.
The analytical first order and total sensitivity indices were derived in \shortcite{Homma1996}
and are given in Table \ref{sensitivity_ishigami}.
The values indicate a vanishing first order influence of $X_3$
but a significant coupling term of $X_1$ and $X_3$ as indicated in the total effect indices.

In our analysis again the subset averaging and the matrix combination approach
are applied using 10.000 Latin Hypercube samples. The regression based approach is not used due to the highly nonlinear functional behavior, 
which can not be represented by low order polynomials.
The results of the analyses assuming uncorrelated and even correlated inputs are given additionally in Table \ref{sensitivity_ishigami}.

The correlation has been assumed in the second case between $X_1$ and $X_3$ as $\rho_{23} = 0.5$. In the Nataf model this correlation coefficient is 
transformed to $\tilde \rho_{23} = 0.518$ in the correlated Gaussian space using the iteration in Eq.~(\ref{nataf}).
In the third case additional correlations are introduced ($\rho_{12} = 0.3$, $\rho_{23} = 0.8$, $\tilde \rho_{12} = 0.313$, $\tilde \rho_{23} = 0.814$).
\begin{table}[th]
\tiny
\begin{tabular}{lccccccccccccc}
    \hline
	& \multicolumn{2}{c}{Analytical} && Subset && \multicolumn{4}{c}{Matrix combination}\\
	 & $S_i$ & $S_{T_i}$ && $\hat S_i$ &&  $\tilde S_i^C$ & $\tilde S_{T_i}^C$ & $\tilde S_i^U$ & $\tilde S_{T_i}^U$ \\
   \hline
	$\rho_{23} = 0.0$ \rule[0mm]{0pt}{2.5ex}\\
   $X_1$ & 0.314  & 0.557 &&  0.324   &&	 0.330 & 0.570 &	 0.330 & 0.570\\
   $X_2$ & 0.442  & 0.442 &&  0.461   &&	 0.456 & 0.429 &	 0.456 & 0.429\\
   $X_3$ & 0.000  & 0.244 &&  0.008   &&	 0.008 & 0.251 &	 0.008 & 0.251\\
   \hline
	\multicolumn{5}{l}{$\rho_{13} = 0.5$} \rule[0mm]{0pt}{2.5ex}\\
   $X_1$ & - & -	&& 0.305   &&	0.312 & 0.453	  & 0.065 & 0.346\\
   $X_2$ & - & -	&& 0.479   &&	0.484 & 0.473	  & 0.484 & 0.473\\
   $X_3$ & - & -	&& 0.174   &&	0.172 & 0.444	  & 0.071 & 0.200\\
   \hline
	\multicolumn{5}{l}{$\rho_{12} = 0.3$, $\rho_{13} = 0.5$, $\rho_{23} = 0.8$} \rule[0mm]{0pt}{2.5ex}\\
   $X_1$ & - & -	&& 0.310   &&	0.306 & 0.799	  & 0.057 & 0.354\\
   $X_2$ & - & -	&& 0.576   &&	0.559 & 0.785	  & 0.022 & 0.454\\
   $X_3$ & - & -	&& 0.213   &&	0.198 & 0.927	  & 0.022 & 0.117\\
   \hline
\end{tabular}
\caption{Sensitivity indices of the Ishigami test function assuming uncorrelated and correlated inputs}
\label{sensitivity_ishigami}
\end{table}
For the uncorrelated case the indices obtained with the different methods agree very well with the theoretical values.
Interestingly for the first correlated case, the first order index
of $X_3$ increases significantly due to the correlation.
			
\subsection{Conclusions}
In the presented paper an extension of the classical Sobol' estimator for correlated input variables has
been proposed. In this method the matrix recombination is modified by changing not only the
variable itself but also its coupling terms with other variables due to the correlations.
For this purpose a decomposition of the sampling matrices in an uncorrelated and a correlated
part is proposed. This decomposition is based on a linear correlation model between the input variables.
Additionally a regression based approach is presented, which is much more efficient, if the model behavior can be represented by the regression basis.
However, attention is required in both methods
in order to perform the decomposition. For non-normally distributed inputs a transformation to the Gaussian space is
necessary before decomposing the samples.
\appendix
In the standard normal space the correlated part of a single variable $Z_i$ can be represented by the remaining variable set 
$\tilde \mathbf{Z}=\mathbf{Z}_{\sim i}$ as follows
\begin{equation}
Z_i^{C,\tilde \mathbf{Z}} = \beta_{Z_i,0} + \sum_{j=1}^{m-1} \beta_{Z_i,j} \tilde Z_j.
\end{equation}
The intercept $\beta_{Z_i,0}$ is zero, since all variables are normally distributed with zero mean.
An estimate of the coefficients $\boldsymbol{\beta}_{Z_i}$
can be obtained from a discrete sample matrix $\boldsymbol{\mathcal{Z}}$ by using linear regression
\begin{equation}
\boldsymbol{\hat \beta}_{Z_i} = (\mathbf{P}_{\tilde \mathbf{Z}}^T\mathbf{P}_{\tilde \mathbf{Z}})^{-1}
\mathbf{P}_{\tilde \mathbf{Z}}^T \boldsymbol{\mathcal{Z}}_{(i)},
\end{equation}
where $\boldsymbol{\mathcal{Z}}_{(i)}$ is the $i$th column of $\boldsymbol{\mathcal{Z}}$ and the level matrix $\mathbf{P}_{\tilde \mathbf{Z}}$ of the reduced variable
set contains all columns of $\boldsymbol{\mathcal{Z}}$ without the $i$th column.
The entries of $\mathbf{P}_{\tilde \mathbf{Z}}^T\mathbf{P}_{\tilde \mathbf{Z}}$ can be obtained from the reduced sample matrix $\boldsymbol{\tilde \mathcal{Z}}$ 
as follows
\begin{equation}
(\mathbf{P}_{\tilde \mathbf{Z}}^T\mathbf{P}_{\tilde \mathbf{Z}})_{kl} = \sum_{t=1}^n \tilde \mathcal{Z}_{tk}\tilde \mathcal{Z}_{tl}.
\label{entries_PTP}
\end{equation}
In the case of standard normal variables, the sum in Eq.~(\ref{entries_PTP}) can be formulated in terms of the estimator of the correlation coefficients as follows
\begin{equation}
(\mathbf{P}_{\tilde \mathbf{Z}}^T\mathbf{P}_{\tilde \mathbf{Z}})_{kl} = (n-1) \hat \rho(\tilde Z_k, \tilde Z_l).
\end{equation}
Analogously, the entries of $\mathbf{P}_{\tilde \mathbf{Z}}^T \boldsymbol{\mathcal{Z}}_{(i)}$ can be formulated as
\begin{equation}
(\mathbf{P}_{\tilde \mathbf{Z}}^T \boldsymbol{\mathcal{Z}}_{(i)})_k = \sum_{t=1}^n \tilde \mathcal{Z}_{tk} \mathcal{Z}_{ti} = 
(n-1) \hat \rho(\tilde Z_k, Z_i),
\end{equation}
which finally results in the estimates of the regression coefficients
\begin{equation}
\boldsymbol{\hat \beta}_{Z_i} = (\mathbf{\hat C}_{\tilde \mathbf{Z}\tilde \mathbf{Z}})^{-1}\boldsymbol{\hat \rho}_{\tilde \mathbf{Z},Z_i},
\end{equation}
where $\mathbf{C}_{\tilde \mathbf{Z}\tilde \mathbf{Z}}$ is the correlation matrix of the reduced variable set $\tilde \mathbf{Z}$.
In order to generate samples of the full variable set $\mathbf{Z}$ the input correlation values 
have to be defined. If this is the case, the predefined values instead of the estimators
can be used and the regression coefficients $\boldsymbol{\beta}_{Z_i}$ can be exactly determined.

\acknowledgements
The author would like to thank H. Keitel and A. Saltelli for helpful discussions on this topic.


\end{document}